\documentclass[prb,amsmath,amssymb,amsfonts,twocolumn,nofootinbib, longbibliography]{revtex4-1}
\usepackage{bm,graphicx,mathrsfs}
\usepackage{graphicx}
\usepackage{epsfig}
\usepackage{amsmath,bbm}
\usepackage{amsfonts,amssymb}
\usepackage{times}
\usepackage{color}
\usepackage{mleftright}

\renewcommand{\phi}{\varphi}

\newcommand{\cL}{{\cal L}}

\newcommand{\cS}{{\cal S}}

\newcommand{\tr}[1]{{\rm tr}\left({#1}\right)}

\newcommand{\bq}{\begin{eqnarray}}
\newcommand{\eq}{\end{eqnarray}}

\newcommand{\be}{\begin{equation}}
\newcommand{\ee}{\end{equation}}
\newcommand{\bea}{\begin{eqnarray}}
\newcommand{\eea}{\end{eqnarray}}

\def\>{\rangle}
\def\<{\langle}

\newcommand{\ket}[1]{|#1\rangle}
\newcommand{\bra}[1]{\langle#1|}

\newcommand{\half}{\frac{1}{2}}

\definecolor{james}{rgb}{1,.6,0}

\newcommand{\qed}{}

\def\qed{\leavevmode\unskip\penalty9999 \hbox{}\nobreak\hfill
     \quad\hbox{\leavevmode  \hbox to.77778em{%
               \hfil\vrule   \vbox to.675em%
               {\hrule width.6em\vfil\hrule}\vrule\hfil}}
     \par\vskip3pt}
     
\begin{document}

\newtheorem{theorem}{Theorem}
\newtheorem{lemma}[theorem]{Lemma}
\newtheorem{corollary}[theorem]{Corollary}
\newtheorem{proposition}[theorem]{Proposition}
\newtheorem{definition}[theorem]{Definition}
\newtheorem{example}[theorem]{Example}
\newtheorem{conjecture}[theorem]{Conjecture}
\newenvironment{remark}{\vspace{1.5ex}\par\noindent{\it Remark:}}%
    {\hspace*{\fill}$\Box$\vspace{1.5ex}\par}

\title{Topological transport in the steady state of a quantum particle with dissipation} %A stationary topological transition in 1D}

\author{Michael J. Kastoryano$^{1,2}$ and Mark S. Rudner$^{2,3}$}
\affiliation{$^1$Institute for Theoretical Physics, University of Cologne, Germany}
\affiliation{$^2$Niels Bohr International Academy, University of Copenhagen, 2100 Copenhagen, Denmark}
\affiliation{$^3$Center for Quantum Devices, University of Copenhagen, 2100 Copenhagen, Denmark}

\date{\today}
\begin{abstract}
We study topological transport in the steady state of a quantum particle hopping on a one-dimensional lattice in the presence of dissipation.
The model exhibits a rich phase structure, with the average particle velocity in the steady state playing the role of a non-equilibrium order parameter.
Within each phase the average velocity is proportional to a topological winding number and to the inverse of the average time between quantum jumps. 
While the average velocity depends smoothly on system parameters within each phase, nonanalytic behavior arises at phase transition points.
We show that certain types of spatial boundaries between regions where different phases are realized host a number of topological bound states which is equal to the difference between the winding numbers characterizing the phases on the two sides of the boundary. 
These topological bound states are attractors for the dynamics; in cases where the winding number changes by more than one  when crossing the boundary, the subspace of topological bound states forms a dark, decoherence free subspace for the dissipative system.
Finally we discuss how the dynamics we describe can be realized in a simple cavity or circuit QED setup, where the topological boundary mode emerges as a robust coherent state of the light field.
\end{abstract}
\maketitle

%%%%%%%%%%%%%%%%%%%%%%%%%%%%%%%%%%%%%%%%%%%%%%%%%%%%%%%%%%%%%%
%%%%%%%%%%%%%%%%%                              Introduction                            %%%%%%%%%%%%%%%%%%%%%
%%%%%%%%%%%%%%%%%%%%%%%%%%%%%%%%%%%%%%%%%%%%%%%%%%%%%%%%%%%%%%

%\section{Introduction}

Advances in technical capabilities for probing and controlling matter at the quantum level have brought new focus onto the roles of quantum coherence and entanglement in a variety of phenomena occurring both in nature and in ``synthetic'' systems created in the laboratory.
As a result, many fundamental questions about the role of decoherence and dissipation in open system quantum dynamics have become both experimentally addressable and highly relevant for progress in the field.
Moreover, while dissipation has traditionally been viewed as playing an antagonistic role in quantum dynamics, many researchers in the condensed matter \cite{schuetz2013,shankar2013,kimchi2016,fitzpatrick2017}, quantum optics \cite{krauter2011,barreiro2011,diehl2008,muller2012} and quantum information \cite{verstraete2008,vollbrecht2011} communities have begun searching for ways of actively using dissipation to enable and control a variety of quantum phenomena.

Many previous works on dissipative engineering have provided ways of realizing and generalizing phenomena known from studies of closed (non-dissipative) quantum systems.
Engineered dissipation has been proposed as a means to achieve steady states of quantum dynamical systems with interesting properties such as: being the result of a quantum computation~\cite{verstraete2008,kastoryano2013}, hosting special types of entanglement~\cite{kraus2008,kastoryano2011}, exhibiting nontrivial topological features~\cite{bardyn2013,diehl2011top,budich2015}, and realizing ground and Gibbs states of many-body systems~\cite{kastoryano2016}. 
Seminal experiments have already demonstrated dissipative state preparation in small quantum systems~\cite{lin2013,barreiro2010exp,shankar2013,leghtas2013}. 

\begin{figure}
\centering
  \includegraphics[width=0.8\columnwidth]{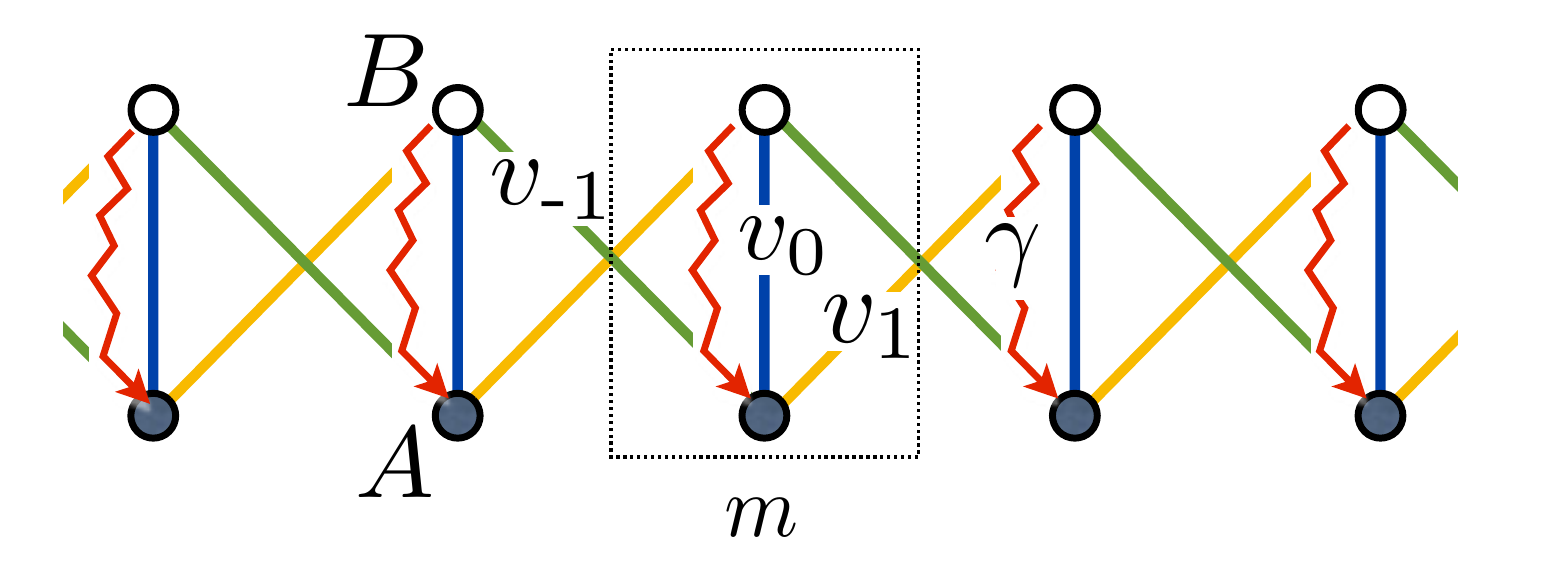}
    \caption{Model for dissipative topological transport, Eqs.~(\ref{eqn:ham}) and (\ref{eqn:master}), restricted to nearest neighbor hopping, $J = 1$. A particle with two internal states $A$ and $B$ hops on a one dimensional lattice with unit cells indexed by the integer $m$. Coherent hopping is indicated with solid lines, while dissipative jumps are shown as wavy arrows. The amplitude for intracell transitions between $A$ and $B$ states is denoted by $v_{ 0}$. Intercell hopping is always accompanied by a transition between $A$ and $B$ states. Amplitudes for hopping to nearest-neighboring unit cells to the left and to the right are denoted by $v_{-1}$ and $v_{1}$, respectively. Intracell dissipative jumps from $B$ to $A$ occur %continuously 
at a rate $\gamma$. } 
    \label{fig0}
\end{figure}

In addition to providing means to realize states or to mimic phenomena known for closed (non-dissipative) systems, dissipation also opens the possibility to explore fundamentally new types of quantum phenomena in which the interplay between coherent and incoherent dynamics gives rise to qualitatively new behavior.
A number recent works have analyzed dynamics in non-Hermitian (including ``$\mathcal{PT}$-symmetric'') lattice systems\cite{rudner2009, longhi2013, liang2013, shen2018,lieu2018}, and investigated novel aspects of topology and bulk-boundary correspondences 
therein\cite{rudner2009, esaki2011, lee2016, rudner2016, leykam2017,klett2017,gong2018top,venuti2017top}.
Here we study a new type of topological transport that occurs in the steady state of a quantum particle hopping on a one dimensional lattice, subjected to (Markovian) dissipation (see Fig.~\ref{fig0}). 
The model can be seen as a particle-conserving relative of the non-Hermitian quantum walks studied in Refs.~\onlinecite{rudner2009,rudner2010,Rapedius2012,rudner2016,zeuner2015,Huang2016, rosenthal2018,harter2018fragile}.
In those works, freely propagating particles (or wave packets) were found to exhibit quantized average displacements before their eventual decay through absorbing sites of the lattice.  
As we will show, the existence of true steady states in the particle conserving problem that we consider gives rise to qualitatively new phenomena.

%%%%%%%%%%%%%%%%%%%%%%%%%%%%%%%%%%%%%%%%%%%%%%%%%%%%%%%%%%
%%%%%%%%%%%%%%%%%%%%%%%%%%%%%%%%%%%%%%%%%%%%%%%%%%%%%%%%%

The system we study can realize a range of dynamical phases, characterized by different steady state transport properties and indexed by an integer-valued topological invariant (winding number).
The steady-state average particle velocity serves as a non-equilibrium order parameter for the system; its value in each phase is proportional to the winding number that characterizes that phase and to the inverse of the average time between quantum jumps induced by the dissipation. 
While the average particle velocity depends smoothly on parameters within each phase, its behavior is nonanalytic at phase transition points where the winding number changes.

For systems with spatially-inhomogeneous hopping coefficients, we find an interesting and unusual form of bulk-boundary correspondence.
For a spatial boundary separating regions characterized by different winding number indices, if the winding number to the left of the boundary is {\it larger than} the winding number to the right of the boundary then the system exhibits a stationary decoherence-free subspace 
spanned by topological bound states that are exponentially localized in the vicinity of the boundary. 
The dimension of the decoherence-free subspace is 
equal to the difference between the winding numbers on the two sides of the boundary. 
The topological bound states (and hence the decoherence free subspace) are attractors for the dynamics.
For boundaries where the winding number {increases} from left to right, the particle generically diffuses away from the boundary and no protected modes are found.

The remainder of the paper is organized as follows.  
In Sec.~\ref{sec:model} we introduce the model and recall some basic properties of (Markovian) dissipative quantum systems. 
Then in Sec.~\ref{sec:topology} we characterize the ``bulk'' topological features of the model, in the translationally-invariant case. 
In Sec.~\ref{sec:bulkedge} we discuss the novel bulk-edge correspondence of the system. 
Finally, in Sec.~\ref{sec:physimp} we discuss a cavity or circuit QED based realization of the phenomena that we describe, and conclude with a discussion in Sec.~\ref{sec:Discussion}.

%%%%%%%%%%%%%%%%%%%%%%%%%%%%%%%%%%%%%%%%%%%%%%%%%%%%%%%%%%%%%%
%%%%%%%%%%%%%%%%%                             Theory                                   %%%%%%%%%%%%%%%%%%%%%
%%%%%%%%%%%%%%%%%%%%%%%%%%%%%%%%%%%%%%%%%%%%%%%%%%%%%%%%%%%%%%

\section{Problem setup}\label{sec:model}

We consider a quantum particle (the ``walker'') with two internal states, $A$ and $B$, coherently hopping on a lattice of $N$ sites with periodic boundary conditions (see Fig.~\ref{fig0}). 
The coherent part of the evolution is described by the  Hamiltonian:
\be 
   H = \sum_m \Delta \ket{B\, m}\bra{B\, m} + \sum_{m,j}[ v_j \ket{B\, m+j}\bra{A\, m}\, +\, h.c.],\! %\sum_{m, n} v^{\alpha\beta}_{n-m} \ket{\alpha,m}\bra{\beta,n},  %+ h.c.\big) %\sum_{j=-J}^J v_{m,j} \ket{A,m}\bra{B,m+j}+h.c.
\label{eqn:ham}
\ee
where $\ket{A\, m}$ and $\ket{B\,m}$ represent states with the particle on site $m$ with internal states $A$ and $B$, respectively, and %, respectively, with $-N/2 \le m < N/2$.
 $\Delta$ is the energy difference between internal states $A$ and $B$. 
We take the hopping to be local with finite range $J$, such that $v_j = 0$ for $|j| > J$. 
Initially we take the system to be translationally invariant; later we will consider situations where the hoppings $v_j$ also depend smoothly on position, $m$. 
Throughout this work we take all hopping coefficients to be real and non-negative.

Note that hopping in our model is always accompanied by a change of the internal state.
This condition corresponds to the ``weak bipartite constraint'' of the associated problem discussed in Ref.~\onlinecite{rudner2016}, and allows a sharp distinction between phases to persist when the hopping varies in space (see discussion in Sec.~\ref{sec:bulkedge} below).
Such a structure arises naturally in certain physical realizations, such as the cavity QED setup in Sec.~\ref{sec:physimp} below, where coherent hopping represents the exchange of excitations between a two level system and an oscillator mode (with photon number $m$).

To investigate dissipative dynamics, we study the time evolution of the reduced density matrix of the system, $\rho$, in the presence of coupling to an environment.
We consider the situation where the $B$ states are metastable, such that coupling to the environment induces spontaneous (incoherent) transitions from $B$ states to $A$ states.
Assuming that the memory time of the environment is short, the equation of motion for $\rho(t)$ is Markovian and takes the Lindblad form $\dot{\rho} = \cL(\rho)$, with
\be 
 \cL(\rho)=-i[H,\rho]+\sum_{m}L_m \rho L_m^\dag-\half\{L^\dag_mL_m,\rho\}_+. \label{eqn:master}
\ee
Here $\{\cdot,\cdot\}_+$ denotes the anti-commutator.

We will discuss the behavior with different choices for the ``jump operators'' $\{L_m\}$ appearing in Eq.~(\ref{eqn:master}).
Physically, these choices differ in the way that coherence between different sites is affected by the dissipation.
For most of the paper we will take $L_m=\sqrt{\gamma}\ket{A\,m}\bra{B\,m}$, where $\gamma > 0$ is the (uniform) decay rate.
With this choice, each quantum jump erases all coherence between different sites (i.e., the environment ``measures'' the position of the particle when it transitions from $B$ to $A$).
We will also discuss the behavior where there is a single jump operator $L = \sqrt{\gamma}\ket{A}\bra{B}\otimes {\bf 1}_m$, where ${\bf 1}_m$ is the identity operator in $m$-space.
This form preserves coherence in the $m$-sector during the transition from $B$ to $A$. 

Unlike the unitary time evolution of closed quantum systems, which preserves the orthogonality between states, the dissipative evolution under Eq.~(\ref{eqn:master}) generally drives all initial states into a lower-dimensional ``stationary subspace'' $\cS$ at long times~\footnote{The stationary subspace is comprised of all states $\rho$ satisfying $\cL(\rho) = 0$.}.
An especially important class of stationary states are \textit{dark states}.
These are states in which a particular pattern of coherence builds up such that the system completely decouples from its environment (and thus becomes immune to further decoherence).
Mathematically, dark states have the special property that they are annihilated by all jump operators, i.e., $L_m\ket{\psi_{\rm dark}}=0$ for all $m$, and are also eigenstates of the Hamiltonian (though not necessarily the ground state), $H\ket{\psi_{\rm dark}} \propto \ket{\psi_{\rm dark}}$. 
If a system has a unique stationary state, such that the dimension of the stationary subspace is 1, then the stationary state is  dark if and only if it is pure~\cite{kraus2008}.

%%%%%%%%%%%%%%%%%%%%%%%%%%%%%%%%%%%

\section{Phase structure of steady states}\label{sec:topology}

In this section, we study the stationary states of the model described in Sec.~\ref{sec:model}, with $L_m = \sqrt{\gamma}\ket{A\,m}\bra{B\,m}$, and map out the resulting phase diagram.
The main result of this section is that the average steady-state velocity $\bar{V}$, which plays the role of an order parameter in our system, is (for $N\rightarrow \infty$): % given by
\be 
  \bar{V} = \frac{a\nu}{\tau}.\label{eqn:topcurrent}
\ee
Here $\tau$ is the average time between dissipative jumps, $\nu$ is an integer-valued topological index, and $a$ is the lattice constant. 
We derive Eq.~(\ref{eqn:topcurrent}) and expressions for $\nu$ and $\tau$ below.

The velocity of the walker is defined through the time derivative of its position, given by the operator $\hat{m} = \sum_{m'\alpha}m'a\,\ket{\alpha\,m'}\bra{\alpha\,m'}$, where $\alpha = \{A,B\}$. 
Using Eq.~(\ref{eqn:master}), and the fact that the jump operators $\{L_m\}$ do not affect the position of the walker, the velocity can be written as $V = -i\langle [H, \hat{m}] \rangle$, where $\langle \mathcal{O} \rangle = {\rm tr}\,(\mathcal{O}\rho)$.

Due to the translation-invariance of the model described by Eqs.~(\ref{eqn:ham}) and (\ref{eqn:master}), its steady states are conveniently analyzed in the Fourier basis spanned by the states $\ket{\alpha\, k} = \frac{1}{\sqrt{N}}\sum_m e^{ikma}\ket{\alpha\,m}$. %, with $\alpha = \{A,B\}$.
We expect  that the steady state achieved at long times is translation-invariant.
As a consequence, the steady state density matrix $\bar{\rho}$ is block diagonal in the Fourier basis: $\bra{\alpha\,k}\bar{\rho}\ket{\beta\,k'} \propto \delta_{kk'}$.
In terms of the $2 \times 2$ block matrices $\bar{\rho}_k$ defined via $(\bar{\rho}_k)_{\alpha\beta} = \bra{\alpha\,k}\bar{\rho}\ket{\beta\,k}$, the steady state velocity [Eq.~(\ref{eqn:topcurrent})] takes the form:
\be 
  \bar{V} = \sum_k \tr{\frac{dH_k}{dk}\bar{\rho}_k}\label{eqn:velocity}.
\ee
Here $H_k$ is the $2 \times 2$ Bloch Hamiltonian associated with the tight-binding problem in Eq.~(\ref{eqn:ham}),
\be
  H_k = \left(\begin{array}{cc}
               0 & c_k \\
             c_k^* & \Delta \end{array}
         \right), \quad c_k = \sum_{j=-J}^Jv_je^{i k ja}. \label{eqn:Hamk}
\ee
In Eq.~(\ref{eqn:Hamk}) we use the ordered basis where the $A$ components come first (i.e., the $AA$ matrix element is the top left entry). 

In order to evaluate $\bar{V}$, we need to solve for the (unique) steady state of Eq.~(\ref{eqn:master}).
The jump operators of the form $\ket{A\,m}\bra{B\,m} = \frac{1}{N}\sum_{kk'}e^{-i (k - k')m a}\ket{A\,k}\bra{B\,k'}$ couple different momentum sectors.
Therefore, finding the steady state $\bar{\rho}$ satisfying $\cL(\bar{\rho}) = 0$ appears to be a complicated, high-dimensional algebraic problem.
However, using the translation-invariance of the steady state, we now show that $\bar{\rho}$ can in fact be constructed analytically from the solutions of a simpler auxiliary problem with the single (momentum-conserving) jump operator $L = \sqrt{\gamma}\ket{A}\bra{B}\otimes {\bf 1}_m$.

When only the single jump operator $L$ is present, each $2 \times 2$ diagonal block $\rho^{(0)}_k$ obeys a separate master equation
\be 
  \dot{\rho}^{(0)}_k = -i\big[H_k,\rho^{(0)}_k\big]+ L\rho^{(0)}_kL^\dag-\half\{L^\dag L,\rho^{(0)}_k\},\label{eqn:masterk}
\ee
where in the Pauli matrix representation we have $L = \sqrt{\gamma}\,\sigma^+$.

The steady state condition $\dot{\rho}^{(0)}_{k}=0$ is then a $2\times 2$ matrix equation, which can be solved analytically for each $k$:
\be 
  \bar{\rho}^{(0)}_{k} = \frac{1}{F_k}\left(\!\begin{array}{cc}
                       |c_k|^2 + \Delta^2 + (\gamma/2)^2  & -c_k(\Delta - i\gamma/2)\\ 
                       -c^*_k(\Delta + i\gamma/2)& |c_k|^2 \end{array}
                    \!\right),\label{eqn:rhoss}
\ee
where $F_k$ is a normalization constant. 
For the calculation that follows, it is convenient to take $F_k=2|c_k|^2+\Delta^2 + (\gamma/2)^2$, which yields ${\rm tr} \,\bar{\rho}_k = 1$. 

We now use the solutions $\bar{\rho}^{(0)}_k$ in Eq.~(\ref{eqn:rhoss}) to build up the steady state of the original problem (\ref{eqn:master}) with the site-specific jump operators $\{L_m\}$.
As mentioned above, problem (\ref{eqn:master}) differs from the momentum-conserving problem (\ref{eqn:masterk}) by the fact that the jump operators $\{L_m\}$ couple different momentum sectors.
Crucially, in the steady state, the population within each momentum sector must be constant in time.
Equivalently, the rate of population leaking out of each momentum sector must be exactly compensated by the population scattered in from all other sectors.
Hence, in the steady state, the behavior within each momentum sector is effectively the same as if the scattered population were simply recycled within the sector, as in the momentum conserving problem (\ref{eqn:masterk}).
This observation motivates an ansatz for the steady state density matrix, expressed in terms of the solutions $\bar{\rho}^{(0)}_k$:
\be 
  \bar{\rho} = \sum_k \bar{\rho}_k \otimes \ket{k}\bra{k}, \quad \bar{\rho}_k = p_k\,\bar{\rho}^{(0)}_k,
  \label{eqn:rssint}
\ee
where the populations $p_k$ satisfy $\sum_k p_k = 1$.

To solve for the populations $p_k$ in Eq.~(\ref{eqn:rssint}), we note that each quantum jump caused by any of the operators $L_m$ spreads the walker's momentum uniformly across the Brillouin zone:
\be 
  \sum_m L_m \ket{B\,k}\bra{B\,k}L^\dag_m = \gamma \sum_{k'} \ket{A\,k'}\bra{A\,k'}.
\ee
Thus, independent of the distribution $\{p_k\}$, the rate of particles being scattered {\it into} the different momentum sectors is always {\it uniform in $k$}.
According to the stationarity condition above, the rate of particles scattering {\it out} of each sector must also be uniform in $k$.
Noting that spontaneous emission occurs from state $B$, the out-scattering rate from sector $k$ is given by $\Gamma_k = \gamma\, p_k(\bar{\rho}^{(0)}_k)_{BB}$.
We therefore use Eqs.~(\ref{eqn:rhoss}) and (\ref{eqn:rssint}) to fix $p_k$ by demanding that $p_k(\bar{\rho}^{(0)}_k)_{BB}$ is independent of $k$:
\be 
  \bar{\rho}_k = \frac{1}{Z}
    \left(\begin{array}{cc}
      1+\frac{\Delta^2 + (\gamma/2)^2}{|c_k|^2} & \frac{-\Delta + i\gamma/2}{{c}^*_k} \\
      \frac{-\Delta - i\gamma/2}{c_k}& 1 
    \end{array}\right).\label{eqn:rssk}
\ee
The normalization factor $Z$, enforcing ${\rm tr}(\bar{\rho}) = 1$, is given by 
\be 
  Z = \sum_k\left[2+\frac{\Delta^2 + (\gamma/2)^2}{|c_k|^2}\right].
  \label{eqn:Z}
\ee 
Note that as long as $Z$ is finite, i.e.,  $c_k \neq 0$ for all $k$, there is only one set of $\{p_k\}$ satisfying Eq.~(\ref{eqn:rssk}); this confirms that the steady state is unique. 

Equipped with an analytic expression for the stationary state, we now calculate the steady state velocity, Eq.~(\ref{eqn:velocity}). 
Using Eqs.~(\ref{eqn:Hamk}) and (\ref{eqn:rssk}), we obtain 
\be 
  \bar{V} = \frac{1}{Z} \sum_k \left(\gamma\, {\rm Im}\left[\frac{d}{dk}\log c_k\right]-\Delta {\rm Re}\left[\frac{d}{dk}\log c_k\right]\right)
\label{eqn:windingVelocity}.
\ee

To identify the underlying topological character, we now take the system size to infinity. 
Converting sums over $k$ to integrals via $\sum_k \rightarrow \frac{Na}{2\pi}\oint dk$, we get 
\be 
  \bar{V} = \frac{a\gamma}{\tilde{Z}} \oint \frac{dk}{2 \pi}~ {\rm Im}\left[\frac{d}{dk}\log c_k\right],
  \label{eqn:winding}
\ee
with $\tilde{Z}\!=\! a\oint \frac{dk}{2\pi}\big(2+\frac{\Delta^2\! +\! (\gamma/2)^2}{|c_k|^2}\big)$.  
The integral of ${\rm Re}\,[\frac{d}{dk}\log c_k]$ is zero, since the integrand is an odd function of $k$. 

Remarkably, Eq.~(\ref{eqn:winding}) reveals that the steady state velocity of the walker 
is proportional to a topological winding number: 
\be 
  \nu = \oint \frac{dk}{2\pi}~ {\rm Im}\left[\frac{d}{dk}\log c_k\right]\label{eqn:windingNr}.
\ee
The index $\nu$ in Eq.~(\ref{eqn:windingNr})  takes only integer values, corresponding to the number of non-trivial cycles that the complex function $c_k$ makes around the origin when $k$ varies from  $0$ to $2\pi$. 
As a result, the model described by Eqs.~(\ref{eqn:ham}) and (\ref{eqn:master}) possesses a rich phase diagram, with steady states organized into distinct topological sectors.

The pre-factor $\gamma/\tilde{Z} \equiv 1/\tau$ is the average time between dissipative jumps at stationarity. 
To see this, note that a jump occurs with rate $\gamma$ whenever the walker sits on a $B$ site. 
The probability for a jump to occur per unit time is therefore proportional to the probability of being at a $B$ site, which is precisely ${1/\tilde{Z}}$.
This gives: 
\be 
  \tau =  \frac{a}{\gamma}\oint \frac{dk}{2\pi}~\left(2+\frac{\Delta^2 + (\gamma/2)^2}{|c_k|^2}\right).
  \label{eqn:intpoles}
\ee

Importantly, the average time between jumps, $\tau$, diverges at the critical points separating different topological phases. 
This follows from the fact that the winding number of $c_k$ can only change when $c_{k_*} = 0$ for some $k_*$; when $c_k$ vanishes, however, the steady state becomes dark and the rate of quantum jumps goes to zero. 
To see why the steady state is dark in this case, note that when $c_{k} = 0$, the state $\ket{A\,k}$ is an eigenstate of $H_{k}$ in Eq.~(\ref{eqn:Hamk}), and is annihilated by all $L_m$. 

At a generic phase transition point, $c_k$ will vanish at a single value of $k$.
In this case there is a single dark state and a unique steady state.
If, for a given set of parameters, $c_k$ vanishes at more than one value of $k$, the system will possess a dark stationary subspace with dimension equal to the number of zeros of $c_k$.
Away from phase transition points, the steady state is never  dark.

%%%%%%%%%%%%%%%%%%%%%%%%%%%%%%%%%%%%%%%%%%%%%%

\subsection{Example: dissipative walk with nearest-neighbor hopping}\label{sec:ex1}

%%%%%%%%%%%%%%%%%%%%%%%%%%%%%%%%%%%%%%%%%%%%%%
\begin{figure}
\centering
  \includegraphics[width=\columnwidth]{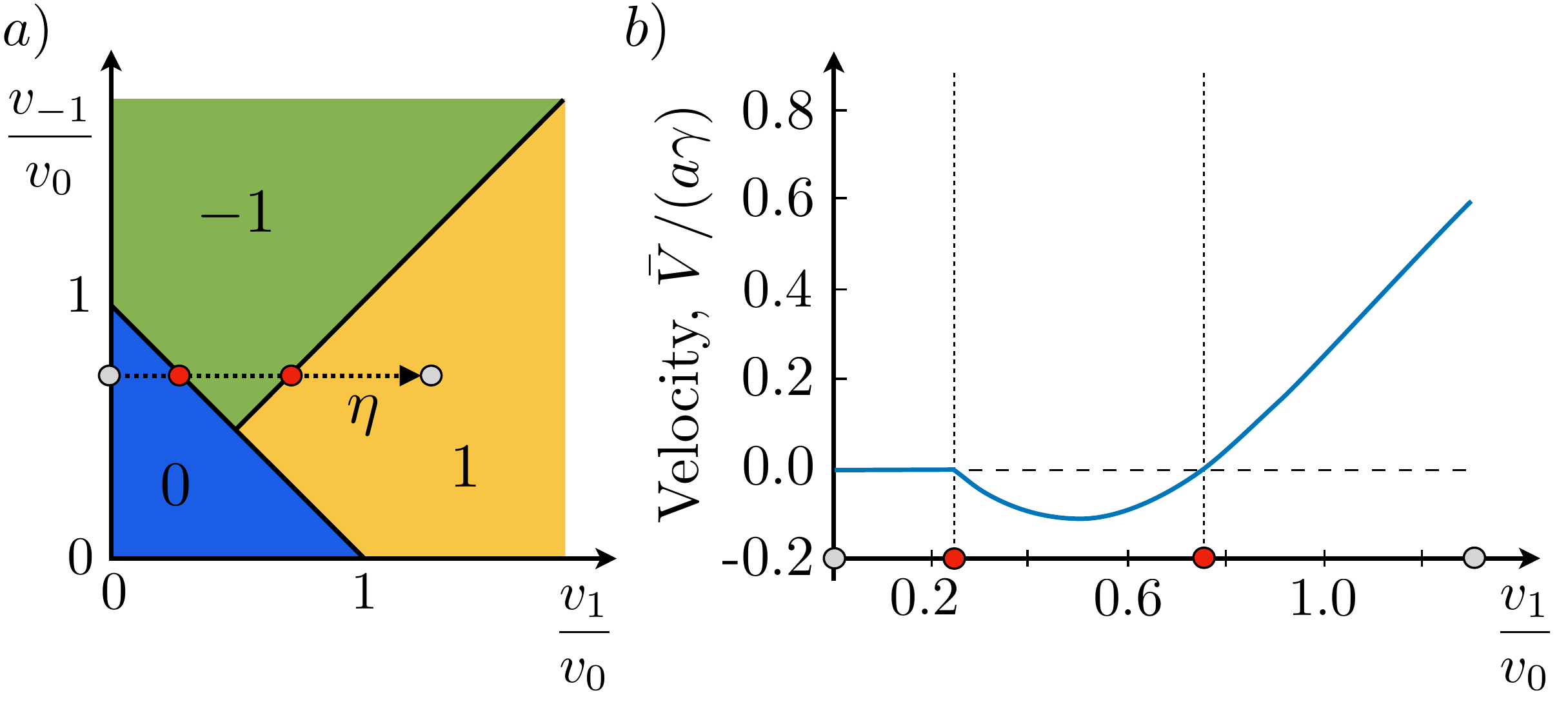}
    \caption{a) Phase diagram of the translationally invariant dissipative walk with  nearest neighbor hopping. Each phase is associated with a value of the winding number, Eq.~(\ref{eqn:windingNr}), which in this example may take the values $\nu=\{-1,0,1\}$. 
b) Stationary particle velocity $\bar{V}$ [Eq.~(\ref{eqn:winding})] along the path $\eta$ shown in panel a), with $v_{-1}/v_0=.75$ and $0\leq v_{1}/v_0\leq 1.25$. Here we take $v_0 = 1$, $\Delta = 1$, and  $\gamma=2$. 
      Along  path $\eta$, the system is initially in a phase with zero average velocity in the steady state. A phase transition is encountered at $v_1/v_0 = 0.25$, beyond which the walker obtains a negative average velocity (i.e., it moves to the left). For $v_1/v_0 > 0.75$, i.e., $v_1 > v_{-1}$, a second transition is passed and the system enters a phase where the walker exhibits a net right-moving steady state motion. 
The average velocity vanishes at the topological transition points. }
    \label{fig1}
\end{figure}
%%%%%%%%%%%%%%%%%%%%%%%%%%%%%%%%%%%%%%%%%%%%%%
We illustrate the results above on an example with maximal hopping range $J = 1$, i.e., $v_j = 0$ for $|j| > 1$. 
We identify three topologically distinct sectors in this nearest-neighbor hopping model. 
By direct evaluation of Eq.~(\ref{eqn:windingNr}) with $c_k = v_0 + v_1 e^{i k a} + v_{-1} e^{-i k a}$, we obtain the winding number $\nu=0$ for $v_0 > v_1 + v_{-1}$, $\nu=-1$ for $v_0 < v_1 + v_{-1}$ with $v_1 < v_{-1}$, and $\nu=1$ for $v_0 < v_1 + v_{-1}$ with $v_1 > v_{-1}$. The corresponding phase diagram is shown in Fig.~\ref{fig1}a. 

The dwell time between quantum jumps, $\tau = \frac{1}{\gamma}\tilde{Z}$, and hence the velocity $\bar{V} = a\nu/\tau$, can also be evaluated analytically.
We do so by writing $c_k$ as a function of the complex variable $z = e^{ika}$: $c(z) = v_0 + v_1 z + v_{-1}/z$.
We then transform the integral over $k$ to an integral over $z$ on the unit circle in the complex plane.
The resulting integral is simple to evaluate via the residue theorem, yielding:

\be 
  \bar{V}=\begin{cases}
    0, & \text{if $v_0 > v_1+v_{-1}$}\\
    \frac{a\gamma B}{(\gamma/2)^2+\Delta^2-B}, & \text{if $v_0<v_1+v_{-1}$ and $v_1< v_{-1}$}\\
   \frac{a\gamma B}{(\gamma/2)^2+\Delta^2+B}, & \text{if $v_0<v_1+v_{-1}$ and $v_1> v_{-1}$},\nonumber
  \end{cases}
\ee
where $B=(v_1-v_{-1})[(v_1+v_{-1})^2-v_0^2]/(v_1+v_{-1})$. The dwell time diverges in the vicinity of phase boundaries, as can be seen explicitly in the above model; $B\rightarrow 0$ at phase boundaries.  

In Fig.~\ref{fig1}b we show a representative trace of the steady state velocity along the path $\eta$ marked by the dotted line in Fig.~\ref{fig1}a.
Along this path, the system crosses three distinct topological sectors, passing from zero, to negative, to positive average velocity. 
At the $\nu=0\rightarrow 1$ transition the velocity exhibits a kink; for the $\nu=-1\rightarrow 1$ transition, the velocity and its derivative are continuous, but there is a discontinuity in the second derivative. 
We observe that the steady state velocity typically exhibits a discontinuity at some finite order of derivative at the topological transition points. However, we have not found a simple relation between the degree of the discontinuity and the change in topological winding number across the transition.

%%%%%%%%%%%%%%%%%%%%%%%%%%%%%%%%%%%

\section{Topological stationary edge modes}\label{sec:bulkedge}
In this section we describe the unusual bulk-boundary correspondence featured by the model defined in Eq.~(\ref{eqn:master}).
To investigate the walker's behavior near spatial boundaries across which the topological index $\nu$ changes its value, we let all the hopping parameters in Eq.~(\ref{eqn:ham}) depend smoothly on the position (unit cell) index, $m$: $v_j \to v_{m,j}$, such that the hopping terms in $H$ become $H_{\rm hop} = \sum_{m,j}v_{m,j}\ket{B m+j}\bra{A m} + h.c.$.
We will work in the $N \to \infty$ limit of a chain with open boundary conditions, which allows us to consider a single boundary between topological phases. 

We focus on the situation where the coefficients $\{v_{m,j}\}$ give rise to two topologically distinct phases in the asymptotic regions $m \to \pm \infty$: one phase extends out to infinity on the left, with winding number $\nu_L$, and another phase extends out to infinity on the right, with winding number $\nu_R$. 
The region in between, which may include finite segments where other topological phases are realized, is considered the ``boundary region.''
As we show below, for $\Delta \nu \equiv \nu_L - \nu_R > 0$ \textit{there exists a dark, decoherence free subspace (DFS) of dimension at least $\Delta \nu$, spanned by states that are exponentially localized around the boundary region.}  
When present, this topological DFS is an attractor for the dynamics: at long times the walker ends up in the DFS, independent of its initialization\footnote{We heuristically argue that, when a DFS of dark edge states exists, then all stationary states of the system lie within the DFS. Suppose a non-dark (``mixed'') stationary state coexists in a system with one or more dark states. From a quantum trajectories point of view, in the mixed stationary state the system repeatedly undergoes dissipative quantum jumps, upon which it is re-initialized into the $A$ subspace.  Each jump will generically induce a finite overlap with the dark state(s); because the dark state components are completely decoupled from the bath, the probability of finding the system in a dark state can only increase over time. In the long time limit, we thus expect the probability for the state of the system to lie within the DFS to go to 1. Hence, a mixed stationary state cannot coexist with dark states.}.

Before going into the technical details of the derivation, we note that the unusual dependence of the bulk-edge correspondence on the {\it sign} of $\Delta \nu$ can be understood intuitively from the fact that the value of $\nu$ expresses the tendency of the walker to move to the right.  
When $\Delta \nu >0$ (meaning $\nu$ is larger on the left of the boundary than on the right), %decreases across a boundary (going from left to right), 
there is a net tendency for the walker to move into the boundary region, where it may become trapped.
If $\nu$ is larger on the right of the boundary than on the left, the walker tends to run away from the boundary region.

To demonstrate the bulk-boundary correspondence described above, we
explicitly construct (pure) dark, stationary state solutions to the master equation (\ref{eqn:master}) of the inhomogeneous system, which are exponentially localized around the boundary region. 

We begin by assuming that there exists a dark stationary state of the form $\ket{\bar{\psi}}=\sum_m \varphi_m\ket{A\,m}$, where $\sum_m|\varphi_m|^2=1$.  
Under the assumption that the hopping coefficients $\{v_{m,j}\}$ are all real, the amplitudes $\varphi_m$ are also real. 
By construction, $\ket{\bar{\psi}}$ only has support on the $A$ subspace; this constraint ensures that $\ket{\bar{\psi}}$ satisfies the dark state condition: $L_m \ket{\bar{\psi}}=0$ for all $m$.
Hence we only need to verify that $H\ket{\bar{\psi}}=E \ket{\bar{\psi}}$, for $H$ given in Eq.~(\ref{eqn:ham}). 

It is easy to check that, for any values of the parameters in Eq.~(\ref{eqn:ham}), the eigenvalue equation for $\ket{\bar{\psi}}$ can only be satisfied for $E=0$. 
(Recall that the energy of the $A$ states is set as the zero of energy.) 
By explicitly acting the Hamiltonian on $\ket{\bar{\psi}}$ and setting the right hand side to zero,
\be 
   H\ket{\bar{\psi}}=\sum_m\sum_{j=-J}^J v_{m,j}\varphi_{m}\ket{B,m+j}=0, \label{eqn:hamact}
\ee
we find a recurrence relation that must be satisfied for each $m$, for any (bound) dark, stationary state:
\be 
   \sum_{j=-J}^J v_{m-j,j}\varphi_{m-j}=0.\label{eqn:recrel}
\ee
To obtain Eq.~(\ref{eqn:recrel}) from Eq.~(\ref{eqn:hamact}), we enforced that the coefficient of each basis state $\ket{B, m}$ in Eq.~(\ref{eqn:hamact}) must vanish.

We recast the recurrence relation in Eq.~(\ref{eqn:recrel}) in terms of transfer matrices: % a pair of transfer matrix equations,
\be 
  \vec{\varphi}_{[m+1]} = T_m\vec{\varphi}_{[m]},\label{eqn:Tp}
\ee
which we use to propagate the state to the right, and 
\be 
  \vec{\varphi}_{[m]} = S_m\vec{\varphi}_{[m+1]},\label{eqn:Sp}
\ee
which we use to propagate the state to the left. 
Here $\vec{\varphi}_{[m]} = (\varphi_{m+J-1},\cdots,\varphi_{m-J})^T$ is a vector of wave function amplitudes centered around unit cell $m$.

To construct  the $2J \times 2J$ transfer matrix $T_m$ defined by Eq.~(\ref{eqn:Tp}), we first isolate the amplitude $\varphi_{m + J}$ (i.e., the first entry in $\vec{\varphi}_{[m+1]}$) in Eq.~(\ref{eqn:recrel}): $\varphi_{m + J} = -\sum_{j = -J + 1}^J \frac{v_{m-j,j}}{v_{m+J,-J}}\varphi_{m-j}$.
Here we assumed for simplicity that $v_{m+J,-J}$ is nonzero; below we comment on the stability of our results to weak (vanishing vs.~non-vanishing) long-range hoppings.
The remaining entries of $\vec{\varphi}_{[m+1]}$, $\{\varphi_{m+J-1}, \ldots \varphi_{m-J+1}\}$, are simply copies of those in $\vec{\varphi}_{[m]}$, shifted down by one.
Thus we have
\be
\label{eq:T}(T_m)_{n'n} = \left\{
\begin{array}{cl}
-\frac{v_{m + J-n,-J + n}}{v_{m+J,-J}}, & \quad n' = 1\\
\delta_{n', n+1}, & \quad n' > 1.
\end{array}
\right.
\ee
The form of the left transfer matrix $S_m$ defined in Eq.~(\ref{eqn:Sp}) can be obtained similarly, by isolating the amplitude $\varphi_{m-J}$ in Eq.~(\ref{eqn:recrel}) and noting that repeated amplitudes are shifted up in $\vec{\varphi}_{[m]}$ relative to their positions in $\vec{\varphi}_{[m+1]}$:
\be
\label{eq:S}(S_m)_{n'n} = \left\{
\begin{array}{cl}
-\frac{v_{m + J + 1 -n,-J - 1 + n}}{v_{m-J,J}}, & \quad n' = 2J\\
\delta_{n', n-1}, & \quad n' < 2J.
\end{array}
\right.
\ee
Here we also assumed $v_{m-J,J} \neq 0$, see discussion below.

We now use the transfer matrices $T_m$ and $S_m$ to construct the wave functions of dark stationary states (which satisfy $H\ket{\bar{\psi}} = 0$). % over the entire chain.
For simplicity, we consider the specific situation of an infinite chain with a single boundary point separating two topologically distinct sectors with uniform couplings on each side of the boundary: $v_{m,j} = v_j^{(L)}$ for $m \le 0$ and $v_{m,j} = v_j^{(R)}$ for $m > 0$.
The hopping parameters in the left and right regions correspond to winding numbers $\nu_L$ and $\nu_R$, respectively, see Eqs.~(\ref{eqn:Hamk}) and (\ref{eqn:windingNr}).
 
Given a set of $2J$ amplitudes $\vec{\varphi}_{[m=0]} = (\varphi_{J-1}, \cdots , \varphi_{-J})^T$ centered around the boundary at $m = 0$, 
the wave function amplitudes centered around unit cell $n$ to the right of the boundary are found by repeatedly applying Eq.~(\ref{eqn:Tp}): $\vec{\varphi}_{[n]} = T_{n-1}\cdots T_1 T_0 \vec{\varphi}_{[m=0]}$.
Similarly, the wave function amplitudes centered around unit cell $-n$ to the left of the boundary are found using Eq.~(\ref{eqn:Sp}): $\vec{\varphi}_{[-n]} = S_{-n}\cdots S_{-2}S_{-1}\vec{\varphi}_{[m=0]}$.

Crucially, not all choices of the $2J$ free variables in $\vec{\varphi}_{[m=0]}$, and values of the winding numbers $\nu_L$ and $\nu_R$, give rise to physical (normalizable) stationary states of the system. 
In particular, it is necessary that the wave function amplitudes do not grow as $m \to \pm \infty$. 
Therefore we seek the conditions under which the transfer matrices yield wave functions that decay exponentially in the left and right regions; these conditions will then tell us the parameter regimes where topological edge states may be found, and the dimension of the dark DFS at the boundary.

The characters of the wave functions generated by the transfer matrices (exponentially growing or decaying in the asymptotic regions) are controlled by the spectral properties of $T$ and $S$ far from the boundary.
(Here we define $T$ and $S$ without $m$ indices, respectively, as the transfer matrices in the translationally-invariant regions far to the right and to the left of the boundary at $m = 0$.)
Indeed, given a set of amplitudes far from the boundary, $\vec{\phi}_{[m]}$ for $m \gg J$, the amplitudes further out are obtained by applying powers of $T$: $\vec{\phi}_{[m+n]} = T^n\vec{\phi}_{[m]}$.
If $\vec{\phi}_{[m]}$ projects onto eigenvectors of $T$ with eigenvalues of modulus greater than or equal to one, then the solution will be non-decaying and therefore non-normalizable.
Each eigenvalue of $T$ with modulus greater than or equal to one therefore provides one linear constraint on the initial amplitudes.
Similarly, the requirement that the wave function must decay exponentially far to the left of the boundary implies that each eigenvalue of $S$ with modulus greater than or equal to one provides an additional linear constraint on the initial amplitudes.
Thus the dimension $\mathcal{D}$ of the dark edge state subspace is given by $\mathcal{D} = 2J - N_S - N_T$, where $2J$ is the number of free parameters in $\vec{\phi}_{[m=0]}$, and $N_S$ and $N_T$ are the numbers of eigenvalues of $S$ and $T$ with modulus greater than or equal to one, respectively. 
In deriving $\mathcal{D}=2J-N_S-N_T$, we have used the fact that, generically, the linear constraints from $S$ and from $T$ are independent. 
If $\mathcal{D} < 0$, this implies that the number of constraints exceeds the number of degrees of freedom, and, generically, a normalizable dark state will not appear.

To determine the values of $N_T$ and $N_S$, we  analyze the characteristic polynomials of $T$ and $S$: the eigenvalues of $T$ and $S$ are given by the roots of the polynomials $t(x)=\sum_{j=-J}^J v^{(R)}_j x^{J-j}$ and $s(x)=\sum_{j=-J}^J v^{(L)}_j x^{J+j}$, respectively.
Comparing with the definition of $c_k$ in Eq.~(\ref{eqn:Hamk}), we observe that $c^{(L)}_k$ and $c^{(R)}_k$ corresponding to the parameters in the left and right regions, respectively, can be related to the characteristic polynomials as:
\begin{equation}
\label{eq:ckst}c^{(L)}_k = e^{-ikJa}\,s(e^{ika}), \quad c^{(R)}_k = e^{ikJa}\,t(e^{-ika}).
\end{equation}

We now show that the dimension of the dark edge state subspace is given by the difference between the winding numbers encoded by $c^{(L)}_k$ and $c^{(R)}_k$, $\mathcal{D} = \Delta\nu$.
Using Eq.~(\ref{eqn:windingNr}), the difference in winding numbers across the boundary is given by $\Delta\nu = \nu_L - \nu_R = \oint \frac{dk}{2\pi}\, {\rm Im}\left[\frac{d}{dk}\log \left({c_k^{(L)}}/{c_k^{(R)}}\right)\right]$.
Equivalently, writing $c^{(L)}_k$ and $c^{(R)}_k$ in terms of $s$ and $t$ using Eq.~(\ref{eq:ckst}),
\bea 
  \Delta\nu &=& \oint\frac{dk}{2\pi}\, {\rm Im}\left[\frac{d}{dk}\log \left(\frac{s(e^{ika})}{e^{2ikJa}t(e^{-ika})}\right)\right]\label{eqn:windingDiff}.
\eea
Changing variables to $z = e^{ika}$, we see that $\Delta \nu$ is equal to the difference between winding numbers of $s(z)$ and $z^{2J}t(z^{-1})$, taken around the unit circle in the complex plane, $|z| = 1$.
Note that $z^{2J}t(z^{-1})$ and $s(z)$ are polynomials  of order $2J$.

According to Cauchy's argument principle, the winding number of a polynomial $p(z)$ taken around a closed contour is equal to the number of zeros of $p(z)$ inside the contour.
In our case, this precisely means that $\Delta \nu$ is equal to the difference between the numbers of roots of $s(z)$ and  $z^{2J}t(z^{-1})$, with $|z| < 1$.
First, using the fact that $s(z)$ has order $2J$, the number of roots of $s(z)$ with $|z| < 1$ is equal to $2J - N_S$, where $N_S$ was defined above as the number of roots of $s(z)$ outside the unit circle\footnote{Here we assume that no eigenvalue has exactly magnitude one, which would require fine-tuning.}. 
Second, we note that the number of roots of $z^{2J}t(z^{-1})$ inside the unit circle is equal to the number of roots of $t(z)$ {\it outside} the unit circle, $N_T$.
Thus we find $\Delta \nu = 2J - N_S - N_T \equiv \mathcal{D}$: the dimension of the dark edge state subspace is precisely the difference of winding numbers across the boundary (provided that it is positive).

%%%%%%%%%%%%%%%%%%%%%%%%%%%%%%%%%%%%%%%%%%%%%%%%%%
\begin{figure}
\centering
  \includegraphics[width=0.9\columnwidth]{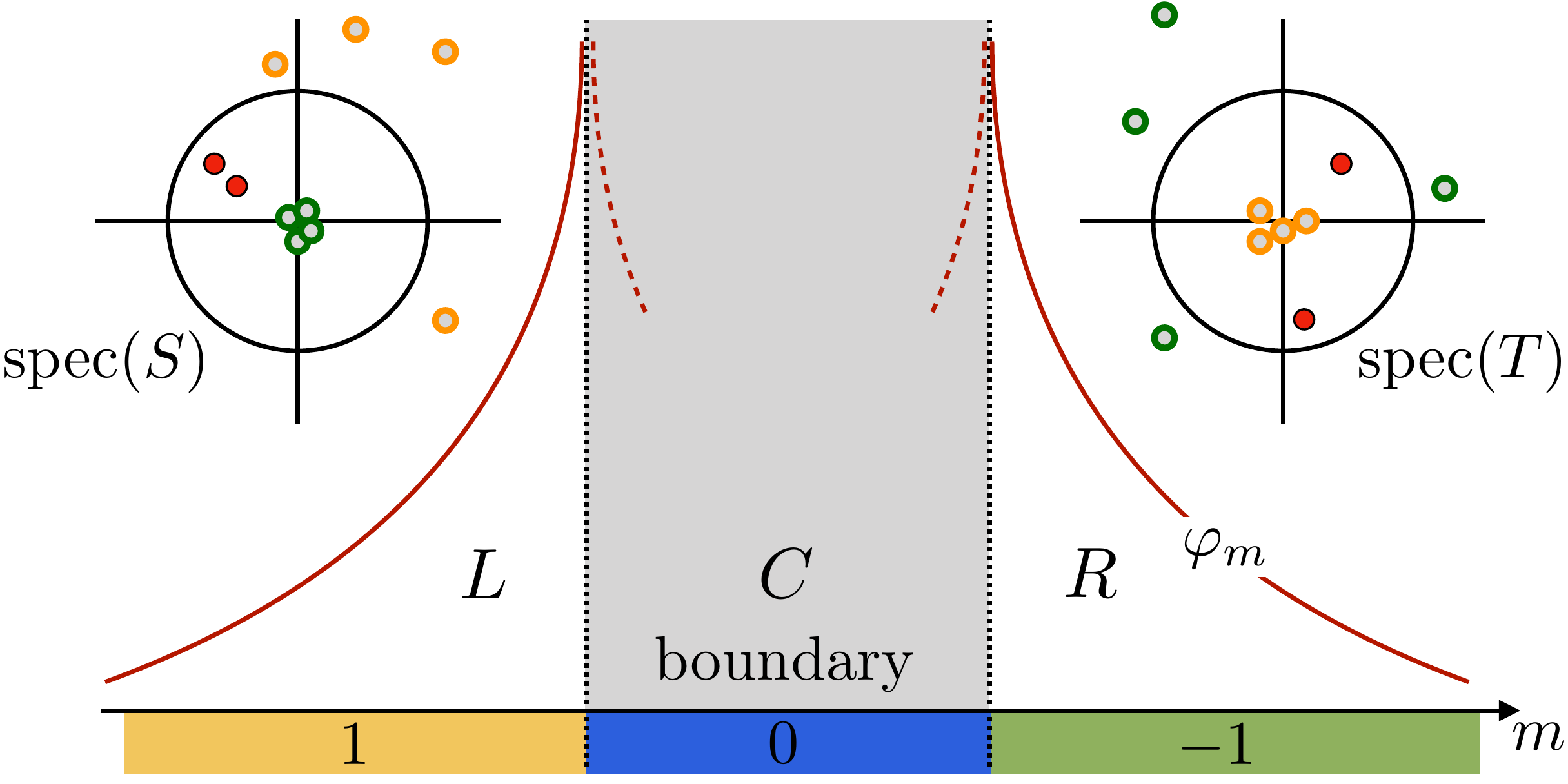}
    \caption{Illustration of stationary dark edge states. The infinite chain is broken into three regions: a finite central region $C$ with $\nu_C=0$ (blue) called the `boundary region,' the left region $L$ with $\nu_L=1$ (yellow) and the right region $R$ with $\nu_R=-1$ (green). Dark edge states are constructed via the right and left transfer matrices $T$ and $S$, Eqs.~(\ref{eq:T}) and (\ref{eq:S}), in regions $R$ and $L$, respectively.  The dimension $\mathcal{D}$ of the dark edge state subspace is given by $2J-N_S-N_T$, where $J$ is the maximal hopping range, and $N_S$ ($N_T$) is the number of eigenvalues of $S$ ($T$) with magnitude larger than one. 
For $\nu_L > \nu_R$, $\mathcal{D} = \nu_L - \nu_R$, while for $\nu_L < \nu_R$ there are no dark states localized near the boundary.
    For  nearest neighbor hopping, $J = 1$, and for $\nu_L = 1$, $\nu_R = -1$, $S$ and $T$ are $2\times2$ matrices with all of their eigenvalues lying inside the unit circle (see insets, red dots). 
Thus $N_S = N_T = 0$ and the system  has a $2J = 2$ dimensional dark subspace. 
If weak uniform hopping of range $J'> J$ is added to the system, the dimensions of $S$ and $T$ increase. However, $\mathcal{D} = 2J' - N'_S - N'_T$ remains unchanged, as the increased value of $2J'$ is compensated by additional eigenvalues of $S$ and/or $T$ that appear outside the unit circle (green and yellow hollow dots).} 
    \label{fig2}
\end{figure}
%%%%%%%%%%%%%%%%%%%%%%%%%%%%%%%%%%%%%%%%%%%%%%%%%%

Before concluding this section, we comment on the stability of the bulk-edge correspondence. 
First, note that our conclusions stem from the properties of $T$ and $S$, the transfer matrices in the asymptotic regions far to the right and to the left of the boundary, respectively.
Therefore the relation between the dimension of the dark state subspace and the topological indices realized on the two sides of the boundary is insensitive to the details of the boundary region. 
Second, one may wonder about the stability of our results to the addition of weak hopping amplitudes of range longer than $J$, which increases the dimensions of the transfer matrices.
Indeed, increasing the dimension of $S$ by (for example) including a nearly vanishing extra hopping amplitude $v_{-J-1} \ll \min v_{j}$ in the left region typically introduces an additional root of $s(z)$ with $|z|$ close to zero (i.e., inside the unit circle).
Correspondingly, introducing a weak hopping amplitude $v_{-J-1}$ in the right region introduces an additional root of $t(z)$ with $|z| \gg 1$, i.e., {\it outside} the unit circle (see Fig.~\ref{fig2}).
In terms of counting constraints, with the increased dimension of $T$ and $S$, one additional piece of initial data is needed to generate a wave function using $T_m$ and $S_m$: the vector $\vec{\varphi}_{[m=0]}$ thus gains one extra component, and has dimension $2J + 1$.
The one additional eigenvalue of $T$ with modulus greater than 1 introduces one extra linear constraint, balancing out the added degree of freedom.
Thus the number of free parameters (and hence the dimension of the dark state subspace) remains invariant when weak longer range hopping amplitudes are introduced.
Moreover, the introduction of the weak additional hopping amplitudes in the left and right regions does not change the winding numbers $\nu_L$ and $\nu_R$, and hence does not change the value of $\Delta \nu$.
Thus the bulk-edge correspondence $\mathcal{D} = \Delta \nu$ is stable to deformations of the hopping amplitudes, including changes to the range of hopping.

We note that the counting arguments above actually give a {\it lower bound} on the dimension of the dark state subspace.
Specifically, we assumed the requirements of avoiding overlap with eigenvectors of $S$ and $T$ with eigenvalues having modulus greater than or equal to one yield {\it independent} constraints on the two sides of the boundary.
Under finely tuned conditions (e.g., where $S$ and $T$ share one or more eigenvectors in common), the constraints may be linearly dependent.
This reduces the number of constraints, and hence increases the dimension of the dark state subspace.
However, such a situation is non-generic, and could be broken by a small deformation of parameters on either side of the boundary.

\subsection{Continued example}

To provide a concrete illustration of the bulk-boundary correspondence described above, we now continue and extend the example of a system with hopping to the nearest neighboring unit cells ($J = 1$), discussed for the bulk in Sec.~\ref{sec:ex1}.
We focus on the situation where the infinite chain is broken up into three regions {$\epsilon = \{L,C,R\}$}, each with constant ($m$-independent) values of the parameters $(v_{m,-1},v_{m,0},v_{m,1})$. 
In a central boundary region $C$ of finite extent {$w$}, $0 \le m < w$, the values of the parameters {$\{v_{C,j}\}$} correspond to the topological sector $\nu_C=0$.
To the left of the boundary region ($\epsilon = L$, $m < 0$) the hopping parameters {$\{v_{L,j}\}$} correspond to the phase $\nu_L=1$, and to the right of the boundary {($\epsilon = R$, $m \ge w$)}, the parameters {$\{v_{R, j}\}$} correspond to $\nu_R=-1$ (see Fig.~\ref{fig2}). 
With these choices of parameters, the walker tends to move toward the boundary region (i.e., it moves to the right in region $L$ and to the left in region $R$).

Assuming that the stationary state is dark ($\ket{\bar{\psi}}=\sum_{m=-\infty}^\infty \varphi_m \ket{A,m}$), we obtain the recurrence relations
\be
   v_{m,0} \varphi_m +v_{m,1} \varphi_{m+1}+v_{m,-1} \varphi_{m-1}=0,\label{eqn:recEx}
\ee
for all $m$.
In explicit form, Eq.~(\ref{eqn:recEx}) along with Eqs.~(\ref{eq:T}) and (\ref{eq:S}) yield the following $2\times2$ transfer matrices: 
\bea 
 \nonumber T_m&=& \left(\begin{array}{cc}
 -{\frac{v_{m,0}}{v_{m+1,-1}}} & -{\frac{v_{m-1,1}}{v_{m+1,-1}}} \\
 1&0\end{array}\right)\\
 S_m &=&\left(\begin{array}{cc}
 0 & 1 \\
 -{\frac{v_{m+1,-1}}{v_{m-1,1}}}& -{\frac{v_{m,0}}{v_{m-1,1}}}
 \end{array}\right).
\eea
Starting with a pair of amplitudes $(\varphi_n,\varphi_{n-1})$, for some integer $n$, we can obtain all of the coefficients $\varphi_m$ with $m<n-1$ by propagating to the left with $\{S_m\}$, and the coefficients with $m>n$ by propagating to the right with $\{T_m\}$. 
In region $L$, where $\nu_L = 1$, the transfer matrix $S$ has both of its eigenvalues {\it inside} the unit circle.
To see this, we take a simple point in the $\nu = 1$ phase: $v_{L,-1} = 0$, $v_{L, 0} = r\bar{v}$, $v_{L,1} = \bar{v}$, where $r < 1$.
In this case $S$ is upper triangular and we can read off the eigenvalues as $0$ and $r$.
Similarly, far to the right of the boundary region the transfer matrix $T$ in region $R$ also has both of its eigenvalues inside the unit circle. (This can be seen, e.g., for the characteristic choice $v_{R,-1} = \bar{v}$, $v_{L, 0} = r'\bar{v}$, $v_{L,1} = 0$, with $r' < 1$, where $T$ has eigenvalues $0$ and $r'$.)
Since all of the eigenvalues of both $T$ and $S$ lie within the unit circle, any choice of $\varphi_n$ and $\varphi_{n+1}$ yields a normalizable wave function that decays exponentially as $m \to \pm \infty$.
Hence the system supports two linearly-independent dark states, $\mathcal{D} = 2$, consistent with the difference in winding numbers across the boundary region.
The walker's probability density will typically be strongly suppressed inside the boundary region if the region is long compared with the decay length of the localized states.
The precise form of the steady state wave function (i.e., which state within the dark state subspace is realized) depends on the initial state of the walker. 

\section{Physical realization}\label{sec:physimp}
In this section we describe a physical setup that maps onto the model described above.
We consider a single two level system with a metastable excited state, interacting with a lossless single mode cavity with resonance angular frequency $\Omega$. The two level system could be, e.g., an atom in a cavity QED setting or a qubit in a superconducting circuit QED implementation. 
Two driving fields are applied: 1) qubit transitions are driven by an ac field of angular frequency $\omega_{\rm a}$, and 2) the qubit-cavity coupling is modulated with angular frequency $\omega_{\rm a-c}$. 
We will work at the two-photon resonance, $\Omega = \omega_{\rm a} + \omega_{\rm a-c}$, such that the qubit and qubit-cavity couplings can mediate a resonant conversion of two driving field photons into one cavity field excitation, and vice-versa.
Within the rotating wave approximation, the coherent part of the system's dynamics is described via a driven version of the Jaynes-Cummings model:
\bea 
H_{\rm JC}(t) &=& \Omega a^\dag a+\half \Delta (1 + \sigma^z)\nonumber\\
&&\ +(\lambda e^{i\omega_{\rm a-c} t}\sigma^+ a + \mu e^{-i \omega_{\rm a} t}\sigma^+ +h.c.),\label{eqn:JC}
\eea
where $a$ is the cavity field annihilation operator, $\sigma^z$ is the Pauli $z$ operator acting in the basis of qubit ground and excited states, $\ket{0}$ and $\ket{1}$, respectively (with $\sigma^z\ket{0} = -\ket{0}$ and $\sigma^z\ket{1} = \ket{1}$), and $\Delta$ is the energy of the  excited state. 
The parameters $\mu$ and $\lambda$ describe the strengths of the (ac) qubit and qubit-cavity couplings. 

Moving to a rotating frame via the transformation $\ket{\tilde{\psi}(t)} = R(t)\ket{\psi(t)}$ with $R(t) = e^{i\Omega t\, a^\dagger a}e^{i\omega_{\rm a}t \frac12(1 + \sigma_z)}$, we obtain a time-independent rotating-frame Hamiltonian $\tilde{H}_{\rm JC} = -i R(t)\frac{d}{dt}R^\dagger(t) + R(t) H_{\rm JC}(t) R^\dagger(t)$: 
\be 
  \tilde{H}_{\rm JC}= \half(\Delta-\omega_{\rm a}) (1 + \sigma^z)+(\lambda \sigma^+ a+\mu \sigma ^+ +h.c.).\label{eqn:JCrf}
\ee
%%%%%%%%%%%%%%%%%%%%%%%%%%%%%%%%%%%%%%%%%%%%%
\begin{figure}
\centering
\includegraphics[width=\columnwidth]{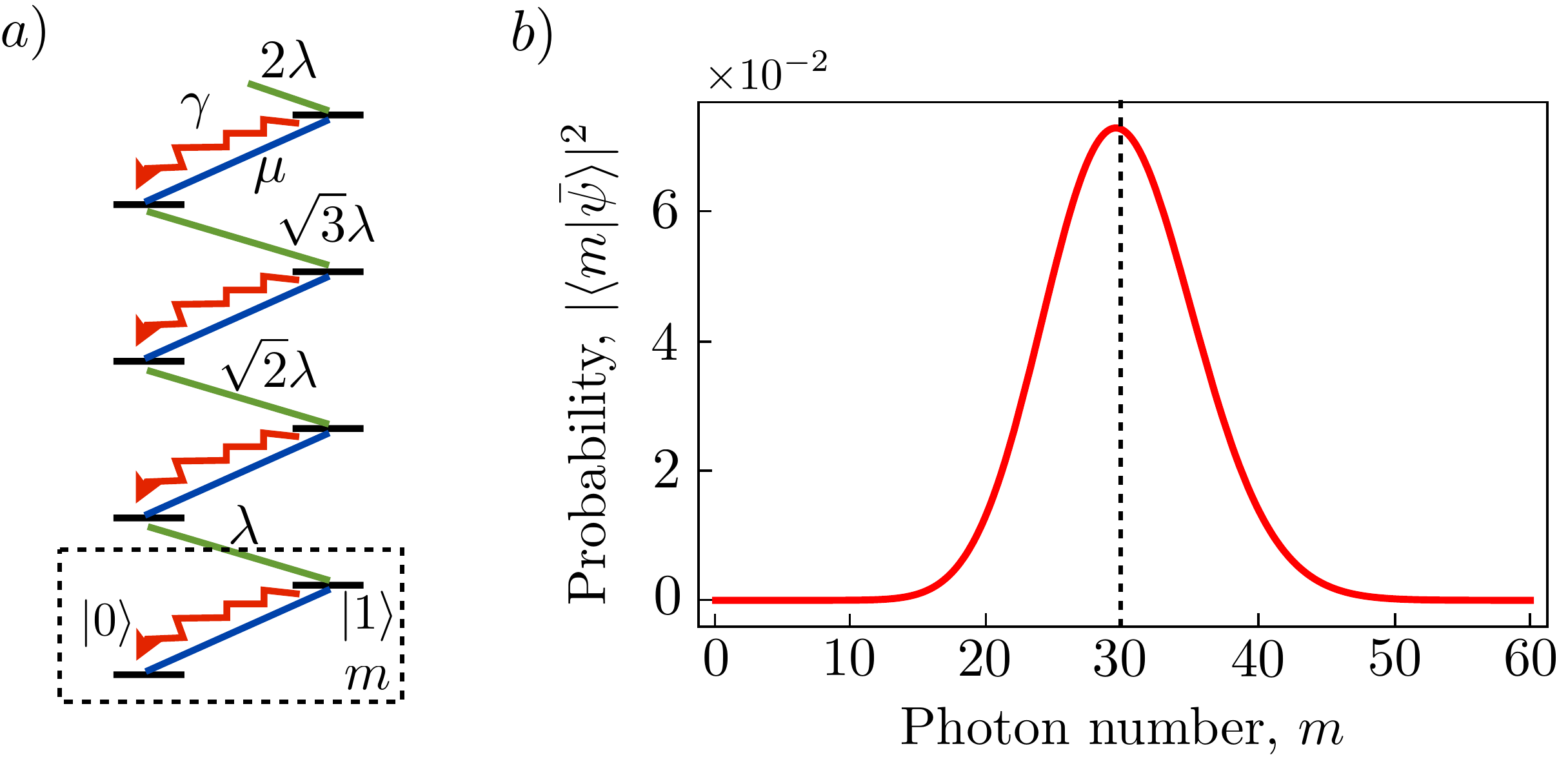}
    \caption{Dynamics of a driven cavity QED system [Eq.~(\ref{eqn:JC})] represented as a dissipative walk through Hilbert space. a) Energy level structure of the cavity QED setup. The effective particle (``walker'') can coherently hop on a semi-infinite lattice of sites corresponding to photon number eigenstates of the cavity field, and indexed by the qubit ground ($\ket{0}$) and excited ($\ket{1}$) states. 
Intracell hopping, in which the cavity photon number $m$ is unaffected, is induced by a driving field of amplitude $\mu$ that acts on the qubit (blue lines).
The driven qubit-cavity coupling allows exchange of excitations between the qubit and cavity (green lines), with an $m$-dependent amplitude $\lambda \sqrt{m}$.
The excited state of the qubit is metastable, and decays at a rate $\gamma$ (represented by wavy arrows). 
b) Steady state site population {$|\langle m |\bar{\psi}\rangle|^2$} [Eq.~(\ref{eqn:coherentstate})] for the cavity QED system with $\mu^2/\lambda^2=30$ (simulated on a finite chain of size $L=60$). 
The population is maximal near $m = \mu^2/\lambda^2$, and is exponentially suppressed away from it. 
The stationary state corresponds to a topological dark edge state, which is localized to the boundary separating regions of differing topological indices. 
}
    \label{figQED}
\end{figure}
%%%%%%%%%%%%%%%%%%%%%%%%%%%%%%%%%%%%%%%%%%%%%

The dynamics generated by $\tilde{H}_{\rm JC}$ in Eq.~(\ref{eqn:JCrf}) can be analyzed in terms of an effective particle hopping between sites of a semi-infinite lattice in Hilbert space, where the unit cells are indexed by the cavity photon number $m$ and each unit cell contains two sites corresponding to the ground and excited qubit states, $\ket{0}$ and $\ket{1}$ (see Fig.~\ref{figQED}a).  
The qubit driving field induces intracell hopping between $\ket{m,\,0}$ and $\ket{m,\,1}$ (with the photon number, $m$, unaffected).
The driven qubit-cavity coupling yields intercell hopping: the amplitude for the effective particle to hop from the state $\ket{m,\,0}$ to the state $\ket{m-1,\, 1}$ in the neighboring unit cell is given by $\lambda \sqrt{m}$, where the $\sqrt{m}$ factor arises from the action of $a$ on the state with $m$ photons.
In the rotating frame, the on-site energy of the excited state is reduced to $\Delta - \omega_{\rm a}$.

We can then map the driven cavity QED problem (\ref{eqn:JC}) to a two-band tight-binding system with spatially-dependent hopping amplitudes.
In the notation of Eq.~(\ref{eqn:ham}), $\ket{0}$ and $\ket{1}$ play the roles of $\ket{A}$ and $\ket{B}$, respectively, and we have $v_{m,0} = \mu$ and $v_{m, -1} = \lambda\sqrt{m}$, with all other hoppings equal to zero.

In addition to the coherent Jaynes-Cummings-type dynamics described above, we assume that the two level qubit may incoherently decay from its excited state to its ground state at a rate $\gamma$. 
The combined coherent and incoherent dynamics are described by the Markovian master equation
\be 
\dot{\rho}=-i[\tilde{H}_{JC},\rho]+\gamma(\sigma^-\rho\sigma^+-\half\{\sigma^+\sigma^-,\rho\}_+).\label{eqn:masterJC}
\ee
The dissipation corresponds to the single jump operator $L=\sqrt{\gamma}\ket{0}\bra{1}\otimes{\bf 1}_m$.
As discussed below Eq.~(\ref{eqn:master}) and in Eq.~(\ref{eqn:masterk}), this type of process preserves coherence of the cavity during decay of the qubit.
In practice, there will always be a finite cavity decay rate $\kappa$; here we assume $\kappa\ll \gamma$, and thus set $\kappa = 0$ for simplicity. 

When $\mu > \lambda$, the system hosts two distinct topological sectors.
In the region $0 \le m < m_c$, with $\sqrt{m_c} = \mu/\lambda$, the winding number $\nu_L$ [Eq.~(\ref{eqn:winding})] is equal to 0 (intracell hopping dominates).
For $m > m_c$, intercell hopping $v_{m,-1}$ dominates and the winding number $\nu_R$ is equal to -1.
(In the right region, the particle tends to move {\it left}, toward the boundary at $m = m_c$.)
Thus we obtain a situation with $\Delta \nu = 1$, and according to the arguments in  Sec.~\ref{sec:bulkedge}  we expect to find a topologically-protected stationary dark state localized around the boundary at $m = m_c$.

By explicit calculation using the recurrence relation (\ref{eqn:recrel}), {$ \mu \varphi_m +  \lambda\sqrt{m+1}\varphi_{m+1} = 0$}, it is straightforward to show that for $\mu>\lambda$ the system has a unique, pure, stationary state: 
\be 
\ket{\bar{\psi}} =e^{-\frac{1}{2}(\mu/\lambda)^2}\sum_{m=0}^\infty(-1)^m\frac{(\mu/\lambda)^m}{\sqrt{m!}}\ket{0,m}.\label{eqn:coherentstate}
\ee
Equation (\ref{eqn:coherentstate}) describes a coherent state of the cavity field, of amplitude $\mu/\lambda$. 
For any initial state, the system tends to this dark state at long times, where it completely decouples from dissipation (assuming $\kappa=0$).

\section{Discussion}\label{sec:Discussion}
In this work we introduced a model of dissipative dynamics in a one-dimensional quantum system, which exhibits a novel and unusual set of dynamical topological  phenomena.
When a spatial boundary is present, separating regions characterized by different topological indices, a topologically-protected decoherence free dark state subspace may form at the boundary.
When such a DFS is present, it is an attractor for the dynamics: independent of the initial state, at long times the state of the system will always belong to the DFS. 
In future work it would be interesting to further explore the nature of this DFS, and its possible applications.

We note that an unusual bulk-boundary correspondence with some analogous features was found for a Hatano-Nelson-type non-Hermitian hopping problem in Ref.~\onlinecite{gong2018top}.
However, the model studied there is of a different nature from the one studied here, in particular lacking particle conservation and a notion of dark states.
A further study of the relation between these problems is an interesting direction for future work.

Constructed as a particle-conserving analogue of the non-Hermitian quantum walk first studied in Refs.~\onlinecite{rudner2009}, the model that we studied can be used to describe a variety of physical systems.
In addition to the cavity/circuit QED based realizations discussed above, following the discussion of Ref.~\onlinecite{rudner2010} we envision that extensions of this model could for example be used to describe nuclear spin pumping in quantum dots.
Whereas the model described in Ref.~\onlinecite{rudner2010} could only be used to assess the short time dynamics of the system, the framework developed in the present work can be used to study the long time dynamics and steady state polarization.

As discussed, e.g., in Refs.~\onlinecite{rudner2016, gong2018top}, topological classification in non-Hermitian systems is a relative concept, dependent on specific physical constraints that are imposed on the system.
In the translationally-invariant case studied here, a natural phase structure emerges based on the condition that the system does not possess any dark states.
In Ref.~\onlinecite{gong2018top}, a wide-ranging classification was developed based on the constraint that the complex spectrum of the system's generator of evolution must avoid one particular value. 
It will be interesting to see what other types of constraints and corresponding topological classifications may naturally emerge in other types of non-Hermitian dynamical systems in the future.

\section{Acknowledgements} We thank S. Diehl, A. S{\o}rensen, and P. Zoller for useful discussions. We acknowledge support from the Villum Foundation and from the DFG (CRC TR 183).

\bibliography{bibDW}

%%%%%%%%%%%%%%%%%%%%%%%%%%%%%%%%%%%%%%%%%%%%%%%%%%%%%%%%%%

\end{document}